\input harvmac
\overfullrule=0pt

%
\def\sqr#1#2{{\vbox{\hrule height.#2pt\hbox{\vrule width
.#2pt height#1pt \kern#1pt\vrule width.#2pt}\hrule height.#2pt}}}

\Title{ \vbox{\baselineskip12pt}}
{\vbox{\centerline{A Representation of Symmetry Generators  }
\bigskip
\centerline{ for the Type IIB Superstring on a }
\bigskip
\centerline{Plane Wave in the U(4) Formalism}}}
\smallskip
\centerline{Gautam Trivedi
\foot{gautamt@physics.unc.edu}
}
\smallskip
\centerline{\it Department of Physics}
\centerline{\it University of North Carolina, Chapel Hill, NC 27599-3255}
\bigskip
\smallskip

\noindent
We calculate the symmetry currents for the type IIB superstring on a maximally supersymmetric plane wave background using the N=(2,2) superconformally covariant U(4) formulation developed by Berkovits, Maldacena and Maoz.  An explicit realization of the U(4) generators together with 16 fermionic generators is obtained in terms of the N=(2,2) worldsheet fields.  Because the action is no longer quadratic, we use a light-cone version to display the currents in terms of the covariant worldsheet variables.

\Date{}

\nref\Maldacena{J.M. Maldacena, ``The Large N Limit of Superconformal Field Theories and Supergravity'', Adv. Theor. Math. Phys. 2 (1998) 231, Int. J. Theor. Phys. 38 (1999) 1113, hep-th/9711200.}

\nref\Kleb{S.S.Gubser, I. Klebanov, A.M. Polyakov,``Gauge Theory Correlators from Noncritical String Theory.''; Phys.Lett.B428:105-114,1998; hep-th/9802109.}

\nref\Figueroa{M. Blau, J. Figueroa-O'Farrill, C. Hull, G. Papadopoulos,``A new maximally supersymmetric background of the IIB superstring theory''; JHEP,0201:047,2002; hep-th/01010242.}

\nref\BMN{D. Berenstein, J. Maldacena, H. Nastase,``Strings in flat space and pp waves from N=4 Super Yang Mills''; JHEP, 0204:013,2002; hep-th/0202021.}

\nref\metsaev{R.R.Metsaev,``Type IIB Green-Schwarz superstring in plane wave Ramond-Ramond Background'';Nucl. Phys. B625,70; hep-th/0112044.}

\nref\MetTsey{R.R. Metsaev, A.A. Tseytlin, ``Exactly solvable model of superstring in plane wave Ramond-Ramond background''; Phys. Rev. D65,126004; hep-th/0202109.}

\nref\Spradlin{M. Spradlin, A. Volovich,``Superstring Interactions in a pp-wave Background''; Phys. Rev. D66, 086004; hep-th/0204146.}

\nref\SpradlinII{M. Spradlin, A. Volovich,``Superstring Interactions in a pp-wave Background II''; JHEP 0301:036,2003; hep-th/0206073.}

\nref\PankI{A. Pankiewicz, B. Stefanski,``PP Wave Light-cone Superstring Field Theory''; hep-th/0210246.}

\nref\PankII{A. Pankiewicz,``More Comments on a Superstring Interaction in a PP-Wave Background;JHEP 0209:056,2002; hep-th/0208209.}

\nref\Schwarz{John Schwarz,``Comments on Superstring Interactions in a Plane-Wave Background''; JHEP 0209:058,2002; hep-th/0208179.}

\nref\SchwarzSprad{Yang-Hui He, John Schwarz, Marcus Spradlin, Anastasia Volovich,``Explicit Formulas for Neumann Coefficients in the Plane Wave Geometry''; hep-th/0211198.}
 
\nref\Huang{Min-xin Huang,``Three point functions of N=4 Super Yang Mills from Light Cone String Field Theory in PP-wave''; Phys. Lett. B542,255,2002; hep-th/0205311.}
 
\nref\Bobkov{K. Bobkov,``Graviton Scalar Interaction in the PP-Wave Background''; hep-th/0303007.}

\nref\BerkTwist{N. Berkovits, ``Calculation of Green-Schwarz Superstring Amplitudes Using the N=2 Twistor Formalism''; Nucl. Phys. B395 (1993) 77; hep-th/9208035.}

\nref\BerkGsTwstNsr{N. Berkovits, ``The Ten-Dimensional Green-Schwarz Superstring is a Twisted Neveu-Schwarz-Ramond String''; Nucl. Phys. B420 (1994) 332; hep-th/9308129.}

\nref\BerkHetr{N. Berkovits, ``The Heterotic Green-Schwarz Superstring on an N=(2,0) Worldsheet''; Nucl. Phys. B379 (1992) 96; hep-th/9201004.}

\nref\BerkMalda{N. Berkovits, J. Maldacena, ``N=2 Superconformal Description of Superstring in Rammond-Rammond Plane Wave Backgrounds''; JHEP, 0210:059,2002;  hep-th/0208092.}

\nref\Maoz{J. Maldacena, L. Maoz,``Strings on pp-Waves and Massive Two Dimensional Field Theories''; JHEP 0204:037,2002; hep-th/0203248.}

\nref\Bonelli{G. Bonelli,``On Type II Strings in Exact Superconformal Nonconstant RR Backgrounds; JHEP 0301:065,2003; hep-th/0301089.}
\newsec{Introduction}
In light of Maldacena's conjecture \refs{\Maldacena-\Kleb}, being able to calculate string correlation functions for strings on curved backgrounds has become very important.  Recently, an exactly solvable, maximally supersymmetric plane wave background of the type IIB superstring was found \refs{\Figueroa}. Berenstein, Maldacena and Nastase \refs{\BMN} have shown the correspondence between strings on this background and SSYM gauge theories. In the light-cone gauge, Metsaev et. al. \refs{\metsaev-\MetTsey} have constructed a worldsheet action for the Type IIB superstring on this plane wave background.  They have also provided a realization of the isometries of the background in terms of their worldsheet fields.  Although this formulation is very useful in calculating the string spectrum, it is not manifestly superconformally invariant thus making it cumbersome for calculating scattering amplitudes.  So far cubic amplitudes for the type IIB superstring on this background have been calculated using the light-cone string field theory formulation\refs{\Spradlin-\Bobkov}.

Another approach to studying the plane wave background has been to use the U(4) formulation developed by Berkovits \refs{\BerkTwist-\BerkHetr}.  In the context of the plane wave background, this formalism has been used by Berkovits and Maldacena \refs{\BerkMalda} to determine consistency conditions for maximally and non-maximally supersymmetric backgrounds \refs{\Maoz}.  They also use it show the exact superconformal invariance of the worldsheet actions on these backgrounds. Similar calculations have been presented for non-constant Ramond-Ramond backgrounds in \refs{\Bonelli}.

The goal of this paper is to obtain an explicit realization of the generators of a U(4) group of background isometries for the plane wave background using the worldsheet fields in the N=(2,2) superconformally covariant U(4) formulation.  In section 2 we begin by reviewing the action for the Type IIB string on a maximally supersymmetric plane wave background and deriving the equations of motion for worldsheet fields.  We then solve the equations in the light-cone gauge in section 3, to obtain an oscillator expansion and impose canonical commutation conditions on the fields and their momenta.  In sections 4 and 5 we construct the U(4) generators in terms of the fields and display their algebra.  We then give a realization of 16 of the 20 manifest fermionic generators in terms of the worldsheet fermions. Finally we discuss the relation between these generators and those constructed by Metsaev et. al. \refs{\metsaev-\MetTsey}.  Although these calculations have been carried out in the light-cone gauge, our generators expressed in terms of the U(4) covariant variables may shed light on how to relax the gauge restrictions on the normal mode expansion and ultimately quantize the U(4) action.

\newsec{Worldsheet action}
As shown by Berkovits and Maldacena \refs{\BerkMalda}, the N=(2,2) superconformally invariant worldsheet action for the type IIB superstring on the maximally supersymmetric plane wave background is

\eqn\action{\eqalign{S_{pp}=\int d^{2}z (&\partial x^{+l} \bar\partial x^{-l}+ 
s^{-l}\bar\partial s^{+l} + \bar{s}^{-l}\partial\bar{s}^{+l} + h^{-l}h^{+l}\cr
+&p^{+}\bar\partial \theta^{-} + p^{-}\bar\partial \theta^{+} 
+ \bar{p}^{+}\partial\bar{\theta}^{-}+\bar{p}^{-}\partial\bar{\theta}^{+}\cr
+& \omega^{+}\bar\partial\lambda^{-} + \omega^{-}\bar\partial \lambda^{+}
+\bar\omega^{+}\partial\bar\lambda^{-}+\bar\omega^{-}\partial\bar\lambda^{+}\cr
+&2[x^{-l}h^{+l}\lambda^{+}\bar\lambda^{+}+x^{+l}h^{-l}\lambda^{-}\bar\lambda^{-}\cr
+&s^{+l}\bar{s}^{+l}\lambda^{+}\bar\lambda^{+}+s^{-l}\bar{s}^{-l}\lambda^{-}\bar\lambda^{-}])}}
\noindent
where $1\leq l,\bar{l}\leq 4$.  This action leads to the following equations of motion:

\eqn\nonlinear{\eqalign{ 
&\bar{\partial}\lambda^{\pm} =0, \qquad \partial\bar{\lambda}^{\pm}=0, \qquad \bar{\partial}\theta^{\pm} =0, \qquad  \qquad \partial\bar{\theta}^{\pm}=0,\cr
&\partial s^{\pm l}+ 2\bar{s}^{\mp l}\lambda^{\mp}\bar{\lambda}^{\mp} = 0,\qquad
\partial \bar{s}^{\pm l}+ 2s^{\mp l}\lambda^{\mp}\bar{\lambda}^{\mp} = 0,\cr
&\partial\bar{\partial}x^{\pm l} +4x^{\pm l}\lambda^{+}\bar{\lambda}^{+}\lambda^{-}\bar{\lambda}^{-} = 0.\cr }}

Even with the $\lambda , \bar{\lambda}, \theta, \bar{\theta}$ restricted to their on-shell values, the rest of the  equations in \nonlinear\ are difficult to solve analytically, since they are not linear, therefore, we first go to the light-cone gauge.  The light-cone gauge conditions are

\eqn\lcguage{\eqalign{&\theta^{\pm}=\bar{\theta}^{\pm}=0,\cr {\rm and \,}
&\lambda^{\pm}=\bar{\lambda}^{\pm}=1.}}

\noindent
which are a particular solution to \nonlinear\ .
Imposing these gauge conditions, the action \action\ reduces to

\eqn\lcaction{\eqalign{S^{l.c.}_{pp}=\int d^{2}z 
&(\partial x^{+l}\bar\partial x^{-l}+ h^{-l}h^{+l}+ 2x^{-l}h^{+l}+2x^{+l}h^{-l}
\cr &+s^{-l}\bar\partial s^{+l}+\bar{s}^{-l}\partial\bar{s}^{+l}+
2s^{+l}\bar{s}^{+l}+2s^{-l}\bar{s}^{-l}).}}

\noindent
Note that this action is similar to the standard light-cone gauge action, but includes the auxilliary fields $h^{\pm l}$. This action leads to solvable equations of motions which are

\eqn\EOM{\eqalign{&\partial\bar\partial x^{\pm l}+4x^{\pm l}=0, \cr
&\bar\partial s^{\pm l} + 2\bar{s}^{\mp l}=0,\cr
&\partial\bar{s}^{\pm l} -2s^{\mp l}=0, }}

\noindent
where the auxilliary fields are set on-shell as $h^{\pm l}=-2x^{\pm l}$. The equations of motion \EOM\ are equivalent to restricting \nonlinear\ to light-cone gauge.  In the next section we will write down the solutions to these equations and thus obtain an normal mode expansion for the worldsheet fields.  Note that the action \action\ is invariant under constant translations of the fermionic worldsheet fields $\theta^{\pm}$ and $\bar{\theta}^{\pm}$. These supersymmetries will be not be manifest in the light-cone gauge. 
 
\newsec{Quantizing the action}
The solutions to \EOM\ for $x^{\pm l}(\sigma,\tau)$ are

\eqn\xplusminus{\eqalign{x^{\pm l}(\sigma,\tau)=&x_{0}^{\pm l}\cos(2\tau)+p_{0}^{\pm l}\sin (2\tau)+\sum_{n\neq 0}{e^{-in\sigma}\over \omega_{n}\sqrt{\pi}}\bigl[ a^{\pm l}_{n}e^{i\omega_{n}\tau}+b^{\pm l}e^{-i\omega_{n}\tau}\bigr]}}

\noindent
where $n \in Z$ and $\omega_{n}=\sqrt{n^{2}+4}$ .The normalization has been chosen for later convenience. The oscillator modes are related to each other by 
the following reality conditions

\eqn\xreal{a^{-l}_{n}=(b^{+l}_{-n})^{\dagger},
\qquad b^{-l}_{n}=(a^{+l}_{-n})^{\dagger}.}

\noindent
The canonical momenta conjugate to $x^{\pm l}(\sigma,\tau)$ are

\eqn\xmom{\eqalign{P^{\mp l}(\sigma,\tau)=-2x_{0}^{\mp l}\sin (2\tau)+2p_{0}^{\mp l}\cos (2\tau)+\sum_{n\neq 0}{ie^{-in\sigma}\over \omega_{n}\sqrt{\pi}}\bigl[ (\omega_{n} \mp n)a^{\mp l}_{n}e^{i\omega_{n}\tau}-(\omega_{n} \pm n)b^{\mp l}_{n}e^{-i\omega_{n}\tau}
\bigr].}}

\noindent
Similarly, solving the equations of motion for the worldsheet fermions leads to
 the following normal mode expansion

\eqn\fermsoln{\eqalign{&s^{\pm l}(\sigma,\tau)=s_{0}^{\pm l}\cos (2\tau)-\bar{s}_{0}^{\mp l}
\sin (2\tau)+ \sum_{n \neq 0}{e^{-in\sigma} \over \sqrt{\pi}\sqrt{4+(\omega_{n}-n)^{2}}}\bigl[s_{n}^{\pm l}e^{i\omega_{n}\tau}-{i \over 2}(\omega_{n}-n)\bar{s}^{\mp l}_{n}e^{-i\omega_{n}\tau}\bigr],
\cr &\bar{s}^{\pm l}(\sigma,\tau)=\bar{s}_{0}^{\pm l}\cos (2\tau)+ s_{0}^{\mp l}\sin (2\tau)
+ \sum_{n \neq 0}{e^{-in\sigma} \over \sqrt{\pi}\sqrt{4+(\omega_{n}-n)^{2}}}\bigl[{-i \over 2}(\omega_{n}-n)s_{n}^{\mp l}e^{i\omega_{n}\tau}+\bar{s}^{\pm l}_{n}e^{-i\omega_{n}\tau}\bigr].}}

\noindent
Equations \xplusminus\ - \xmom\ can be inverted to obtain the oscillator modes

\eqn\bosemodes{\eqalign{x_{0}^{\pm l}&={1 \over 2\pi}\int_{0}^{2\pi} d\sigma \bigl[x^{\pm l}(\sigma,\tau)\cos (2\tau) - {P^{\pm l}(\sigma,\tau) \over 2}\sin (2\tau) \bigr],
\cr p_{0}^{\pm l}&={1 \over 2\pi}\int_{0}^{2\pi} d\sigma \bigl[x^{\pm l}(\sigma,\tau)\sin (2\tau) + {P^{\pm l}(\sigma,\tau) \over 2}\cos (2\tau) \bigr],
\cr a_{n}^{\pm l}&={\pm ie^{-i\omega_{n}\tau} \over 4\sqrt{\pi}}\int_{0}^{2\pi}d\sigma e^{in\sigma}\bigl[(n\mp i\omega_{n})x^{\pm l}(\sigma, \tau) \mp P^{\pm l}(\sigma, \tau)\bigr],
\cr b_{n}^{\pm l}&={\mp ie^{i\omega_{n}\tau} \over 4\sqrt{\pi}}\int_{0}^{2\pi}d\sigma e^{in\sigma}\bigl[(n\pm i\omega_{n})x^{\pm l}(\sigma, \tau) \mp P^{\pm l}(\sigma, \tau)\bigr],}}

\noindent
 Similarly, \fermsoln\ is inverted to obtain the fermionic oscillator modes

\eqn\fermmodes{\eqalign{s_{0}^{\pm l}&= {1 \over 2\pi}\int_{0}^{2\pi}d\sigma\bigl[s^{\pm l}\cos (2\tau) + \bar{s}^{\mp l}\sin (2\tau)\bigr],
\cr \bar{s}_{0}^{\pm l}&= {1 \over 2\pi}\int_{0}^{2\pi}d\sigma\bigl[\bar{s}^{\pm l}\cos (2\tau) - s^{\mp l}\sin (2\tau)\bigr],
\cr s_{n}^{\pm l}&= {2e^{-i\omega_{n}\tau} \over \sqrt{\pi}\sqrt{4 + (\omega_{n}-n)^{2}}}\int_{0}^{2\pi}d\sigma e^{in\sigma}\bigl[s^{\pm l}+ {i \over 2}(\omega_{n}-n)\bar{s}^{\mp l}\bigr],
\cr \bar{s}_{n}^{\pm l}&= {2e^{i\omega_{n}\tau} \over \sqrt{\pi}\sqrt{4 + (\omega_{n}-n)^{2}}}\int_{0}^{2\pi}d\sigma e^{in\sigma}\bigl[\bar{s}^{\pm l}+ {i \over 2}(\omega_{n}-n)s^{\mp l}\bigr].}}
Imposing canonical commutations on the worldsheet fields leads to the nonvanishing commutation relations for the oscillators:

\eqn\comm{\eqalign{&[x_{0}^{\pm l},p_{o}^{\mp k}]=-\delta^{lk}, \qquad[a_{n}^{\pm l},b_{m}^{\mp k}]=\omega_{n}\delta_{n+m}\delta^{lk},
\cr&\{s_{0}^{\pm l},s_{0}^{\mp k}\}=\delta^{lk},\,\, \qquad\{\bar{s}_{0}^{\pm l},\bar{s}_{0}^{\mp k}\}=\delta^{lk},
\cr&\{s_{m}^{\pm l},\bar{s}_{n}^{\pm k}\}=\delta_{m+n}\delta^{lk}.
 }}

\newsec{U(4) Generators}
Having obtained the commutation relations \comm\ , we now give a realization of the U(4) generators and the 16 fermionic generators that are manifest symmetries of the formulation. Note that Berkovits and Maldacena \refs{\BerkMalda} have found 20 manifest fermionic symmetries before going to the light-cone gauge, but as explained above, four of these are not apparent in the light-cone gauge. We begin by defining
\eqn\symj{\eqalign{J^{+l,-k}={1\over 2\pi}\int_{0}^{2\pi} d\sigma : (x^{+l}p^{-k}-x^{-k}p^{+l}+s^{+l}s^{-k}+\bar{s}^{+l}\bar{s}^{-k}):}}
\noindent
where we have used standard rules for normal ordering. Since $1 \leq l,k \leq 4$,\symj\ describes sixteen generators. We now take the symmetric and antisymmetric combinations of these generators to obtain a basis in which an SO(4) subgroup of U(4) is apparent. Consider the following combinations

\eqn\symL{L^{lk}={-i \over 2}(J^{+l,-k}+J^{-l,+k}), \qquad
\bar{L}^{lk}={-i \over 2}(J^{+l,-k}-J^{-l,+k}),}

\noindent
where the generators $L^{lk}$ are antisymmetric under [$l \leftrightarrow k$] and $\bar{L}^{lk}$ are symmetric generators. Using the commutation relations for the oscillators from \comm , we obtain the following algebra for $L, \bar{L}$

\eqn\algebra{\eqalign{&[L^{lk},L^{pq}]=L^{lq}\delta^{kp}-L^{lp}\delta^{kq}-L^{pk}\delta^{lq}+L^{qk}\delta^{lp},\cr
&[\bar{L}^{lk},\bar{L}^{pq}]=L^{lq}\delta^{kp}+L^{lp}\delta^{kq}+L^{kp}\delta^{lq}+L^{kq}\delta^{lp},\cr
&[L^{lk},\bar{L}^{pq}]=\bar{L}^{lq}\delta^{kp}+\bar{L}^{kp}\delta^{lq}-\bar{L}^{lp}\delta^{kq}-\bar{L}^{kq}\delta^{lp}.}} 

\noindent
We can see that $L^{lk}$ are the generators of an SO(4) {\it symmetric} subgroup of U(4).

It can be shown explicitly that the action in the light-cone gauge is invariant under transformations produced by these U(4) generators off-shell.  As an example of this calculation, consider the transformation of the bosonic part of the action under the action of U(4) generators.  We start by noting the transformations of $x^{\pm l}$ under $L^{lk}$ and $\bar{L}^{lk}$ \BerkMalda\ :

\eqn\xtrans{\eqalign{&x^{\pm m}\rightarrow x^{\pm m}+ \Delta_{lk}[L^{lk},x^{\pm m}]
\cr &x^{\pm m}\rightarrow x^{\pm m}+\bar{\Delta}_{lk}[\bar{L}^{lk},x^{\pm m}]
\cr &[L^{lk},x^{\pm m}]= x^{\pm l}\delta^{km}-x^{\pm k}\delta^{lm}
\cr &[\bar{L}^{lk},x^{\pm m}]=\pm \bigl[x^{\pm l}\delta^{km}+x^{\pm k}\delta^{lm}\bigr]}}

\noindent
Where $\Delta, \bar{\Delta}$ are some constants. The bosonic part of the action \lcaction\ with the auxilliary fields $h^{\pm l}$ put on-shell is

\eqn\bosact{\eqalign{S^{l.c.}_{B}&=\int d^{2}z \bigl[\partial x^{+m} \bar{\partial}x^{-m}-4x^{+m}x^{-m}\bigr]}}

\noindent
The variation of the action under the SO(4) symmetric subgroup generators $L^{lk}$ is 

\eqn\svarL{\eqalign{\Delta_{L}S^{l.c.}_{B}&=\int d^{2}z \Delta_{lk} \bigl[ ( \partial x^{+l}\delta^{km}-\partial x^{+k}\delta^{lm})\bar{\partial}x^{-m} +\partial x^{+m}(\bar{\partial}x^{-l}\delta^{km}-\bar{\partial}x^{-k}\delta^{lm})
\cr & -4(x^{+l}\delta^{km}-x^{+k}\delta^{lm})x^{-m} -4x^{+m}(x^{-l}\delta^{km}-x^{-k}\delta^{lm})\bigr]
\cr &=0.}}

\noindent
Furthermore, the variation of the action under the rest of the U(4) generators 
$\bar{L}^{lk}$ is 
\eqn\svarLbar{\eqalign{\Delta_{\bar{L}}S^{l.c.}_{B}&=\int d^{2}z \bar{\Delta}_{lk} \bigl[ ( \partial x^{+l}\delta^{km}+\partial x^{+k}\delta^{lm})\bar{\partial}x^{-m} -\partial x^{+m}(\bar{\partial}x^{-l}\delta^{km}+\bar{\partial}x^{-k}\delta^{lm})
\cr & -4(x^{+l}\delta^{km}+x^{+k}\delta^{lm})x^{-m} +4x^{+m}(x^{-l}\delta^{km}+x^{-k}\delta^{lm})\bigr]
\cr &=0.}}

\noindent
Equations \svarL\ - \svarLbar\ show that the bosonic part of the action is 
manifestly invariant under the U(4) transformations generated by \symL\ . One can similarly demonstrate that the fermionic part of the action is also manifestly invariant under these U(4) transformations. Next, we give a light-cone realization of the sixteen fermionic symmetries whose action on the covariant worldsheet superfields was given in \refs{\BerkMalda}.

\newsec{Fermionic generators}
Under the sixteen fermionic generators restricted to  light-cone gauge, the world sheet fields transform as  \refs{\BerkMalda}

\eqn\fermgenact{\eqalign{s^{\pm l}& \rightarrow s^{\pm l}+\epsilon^{\pm l},
\cr \bar{s}^{\pm l} & \rightarrow \bar{s}^{\pm l}+\bar{\epsilon}^{\pm l},}}

\noindent
where $\epsilon^{\pm l}, \bar{\epsilon}^{\pm l}$ are constant fermions. The generators for these symmetries can by realized in terms of the worldsheet fields $s^{\pm l}, \bar{s}^{\pm l}$ as follows

\eqn\susygen{q^{\pm l}= {1 \over 2\sqrt{\pi}}\int_{0}^{2\pi} d\sigma s^{\pm l},
\qquad \bar{q}^{\pm l}= {1 \over 2\sqrt{\pi}}\int_{0}^{2\pi} d\sigma \bar{s}^{\pm l}.}

\noindent
These generators are symmetries of the equations of motion.  The commutation relations are

\eqn\fermialg{\eqalign{&\{q^{\pm l},\bar{q}^{\pm k}\}=0,\cr
&[L^{lk},q^{\pm m}]= q^{\pm l}\delta^{km}-q^{\pm k}\delta^{lm},\qquad [\bar{L}^{lk},q^{\pm m}]=\pm[q^{\pm l}\delta^{km}+q^{\pm k}\delta^{lm}],\cr
&[L^{lk},\bar{q}^{\pm m}]=\bar{q}^{\pm l}\delta^{km}-\bar{q}^{\pm k}\delta^{lm},\qquad [\bar{L}^{lk},\bar{q}^{\pm m}]=\mp[\bar{q}^{\pm l}\delta^{km}+\bar{q}^{\pm k}\delta^{lm}].}}

After computing the generators, we compared them to the isometries used  by Metsaev \refs{\metsaev} consisting of 30 even and 32 odd generators. We note that the bosonic generators $L^{ij}$ of an SO(4) symmetric subroup of U(4) are linear combinations of the two sets of SO(4) generators displayed by Metsaev \refs{\metsaev}.  The remaining 10 generators of the U(4) group and the 16 odd ones are  not simply related to the generators of \refs{\metsaev}. 

\vskip50pt {\bf Acknowledgements:} We thank Louise Dolan for conversations. 
GT is partially supported by the U.S. Department of Energy, 
Grant No. DE-FG02-97ER-41036/Task A.

\listrefs

\bye